# Distributed Quantum Fiber Magnetometry

*Shai Maayani[1], Christopher Foy[1,2], Dirk R. Englund[1,2], and Yoel Fink[1,2,3*]*

1 Research Laboratory of Electronics (RLE), Massachusetts Institute of Technology, Cambridge, MA 02139 USA

2 Department of Electrical Engineering and Computer Science, Massachusetts Institute of Technology, Cambridge, MA 02139 USA

3 Department of Materials Science and Engineering, Massachusetts Institute of Technology, Cambridge, MA 02139 USA

* Corresponding Author: maayani@mit.edu

**Nitrogen-vacancy (NV) quantum magnetometers offer exceptional sensitivity and long-term stability. However, their use to date in distributed sensing applications, including remote detection of ferrous metals, geophysics, and biosensing, has been limited due to the need to combine optical, RF, and magnetic excitations into a single system. Existing approaches have yielded localized devices but not distributed capabilities. In this study, we report on a continuous system-in-a-fiber architecture that enables distributed magnetic sensing over extended lengths. Key to this realization is a thermally drawn fiber that has hundreds of embedded photodiodes connected in parallel and a hollow optical waveguide that contains a fluid with NV diamonds. This fiber is placed in a larger coaxial cable to deliver the required**

**RF excitation. We realize this distributed quantum sensor in a water-immersible 90-meter-long cable with 102 detection sites. A sensitivity of 63±5 nT Hz-1/2 per site, limited by laser shot noise, was established along a 90 m test section. This fiber architecture opens new possibilities as a robust and scalable platform for distributed quantum sensing technologies.**

**Significance Statement:** Here, we constructed a water-immersible 90-meter-long fully-integrated fiber system that allows distributed quantum magnetic sensing over large distances with a sensitivity of 63 nT Hz-1/2. Applications include remote detection of ferrous metals, geophysics, and biosensing.

Current distributed fiber sensors have high sensitivity to temperature,[1,2] strain,[3–5] and pressure[6–8] -- but not to magnetic fields.[9–11] Here, we add distributed spin-based quantum magnetometry to the sensing capabilities of distributed fiber sensors through the integration of NV ensembles. Recent years have seen the rapid advancement of NV solid-state quantum sensors because of their excellent sensitivity to magnetic fields,[12–15] with sub-nT Hz-1/2 sensitivity in the dc limit,[16,17] high dynamic range vector resolution,[18,19] and remarkable long-term stability.[20,21] Recent advances toward component integration have achieved magnetometry point probes consisting of micro-diamonds on fiber facets[22] or nanodiamonds on tapered fibers.[23] However, practical devices will require compact and stable architectures and -- for some applications -- distributed sensor arrays for magnetometry. Here we introduce an NV-based fiber magnetometer composed of hollow-core silica fiber, drawn with a functional cladding containing the necessary electronics for optical detected magnetic resonance (ODMR) magnetometry. We likewise use the fiber as a delivery mechanism for the pump field,[24] allowing spin polarization and readout, while extending component integration by having the fiber serve as a backbone for photodetection and microwave spin control. The culmination of these properties within our fiber enables distributed magnetometry.

The device fabrication is outlined in Figure 1a. The process consists of two thermal fiber drawings. First, we draw a hollow-core silica fiber (index 1.46) through a high-temperature draw tower (2050 $^0$C), which is subsequently polymer-coated (MyPolymer™ OF-145N, UV-curable index 1.445). The polymer coating functions as cladding for light guidance and mechanically protects the brittle silica shell. Next, we produce a polycarbonate preform with embedded, prefabricated photodiodes (PD) located in prescribed locations along the preform. During the low-temperature thermal draw at 280°C, we unspool the hollow-core fiber together with two copper wires into separate prefabricated channels in the preform, (see Figure 1a). As the preform is thermally drawn, the PDs separate axially while the two wires approach each other until making contact with the PDs.[25] The PDs operate in a parallel circuit. We insert the finished fiber into a notched coaxial cable (RG6/U MOHAWK M71003), which supplies the driving MW fields to all the sensing sites simultaneously (see Fig S6).

By feeding the PC preform at 1 mm/min while drawing at 0.8 meters/min, we produce the completed fiber with the desired cross section of 0.9 x 0.9 mm. Figure 1b shows a close-up of the fiber with one of the PDs visible. The cross-section in Figure 1c shows the fiber and copper wires, which straddle the PD (not visible in this cut) as indicated by the dashed lines. Figure 1d shows the 300 m fiber spool. The complete cable assembly is shown in Figure 1e.

The magnetometry relies on an ensemble of micro-diamonds containing NVs in an oil droplet (Nikon 50 Immersion Oil with an index of 1.518) that can be spatially scanned through the hollow-core fiber. We apply an oleophobic coating (3M Novec™ 1720 Electronic Grade Coating) to prevent adhesion of the micro-diamond-oil droplet to the inner silica wall. The 100-μm inner diameter of the silica fiber allows free movement of the oil drop; the 200-μm outer diameter is chosen to enhance mechanical robustness.

This integrated fiber detection architecture spans decades of dimensional scales and different materials to deliver a distributed magnetometer system. Specifically, this fiber ranges from nanoscale NV centers hosted within micron diamond crystals, with 670-micron silicon PDs, to hundred-meter-scale cable. A cross-section of the completed device is illustrated in Figure 2a. The fiber contains periodically spaced sensing modules (average spacing of 17 cm) to measure the magnetic field magnitude $|\boldsymbol{B}(s)|$ at distance $s$ along the fiber. The 532-nm excitation field required for optical detection of magnetic resonances (ODMR) is guided in the silica shell (see Figure 2b).

When the waveguided pump reaches the high-index oil droplet, it is scattered into the micro-diamond-oil droplet, whose 1-mm length matches that of the embedded PD (Si PD T300P Vishay). The integrated PDs differ from the chalcogenide photoresistive materials which have previously been drawn in fibers by having higher responsivity and bandwidth.[26,27] We scan the NV-micro-diamond droplet with air pressure supplied at the back end of the fiber. This droplet contains several thousand high-fluorescence micro-diamonds (Adamas™). Figure 2c is a photograph of the NV droplet in the fiber. When the droplet is aligned with one of the embedded PDs, the red NV fluorescence is detected from the green pump background using a lock-in technique described below. The PDs are connected in parallel through two fiber-embedded copper wires, using a recently developed self-aligning fiber drawing process.[25] The surrounding coaxial cable drives the NV spin resonance for ODMR. The embedded PDs collect the emitted NV fluorescence, and the generated photocurrent is measured using a centralized readout bus located at the end of the fiber.

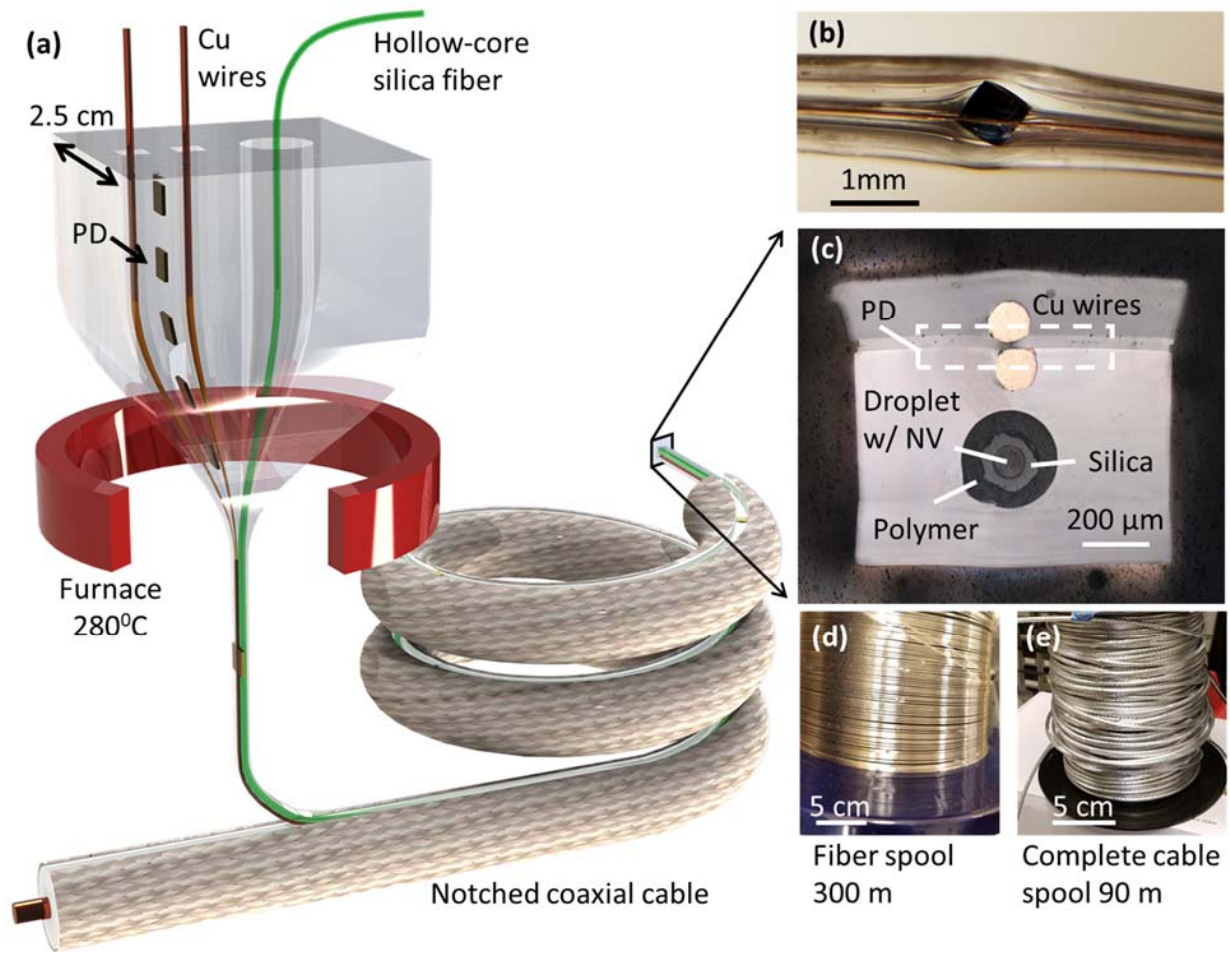

**Figure 1: Fiber fabrication.** **(a)** Illustration of the fiber preform structure and fiber fabrication process. We embed a series of tightly packed PD chips in a polycarbonate (PC) cladding. During our thermal draw, we insert two copper wires, and a hollow-core silica fiber, prepared in advance, through holes in the PC. To achieve the required dimensions of the final device, we control the feed and draw speed of the PC. The cable assembly is completed by inserting the fiber into a notched coaxial cable. **(b)** Close-up of a small section of the fiber with an embedded diode shown. **(c)** Micrograph of the cross-section of the fiber. The bright circles are the inserted copper wires. The dashed box represents the location of embedded diodes. **(d)** The 300 m spool of fiber. **(e)** The completed cable spool (90 m).

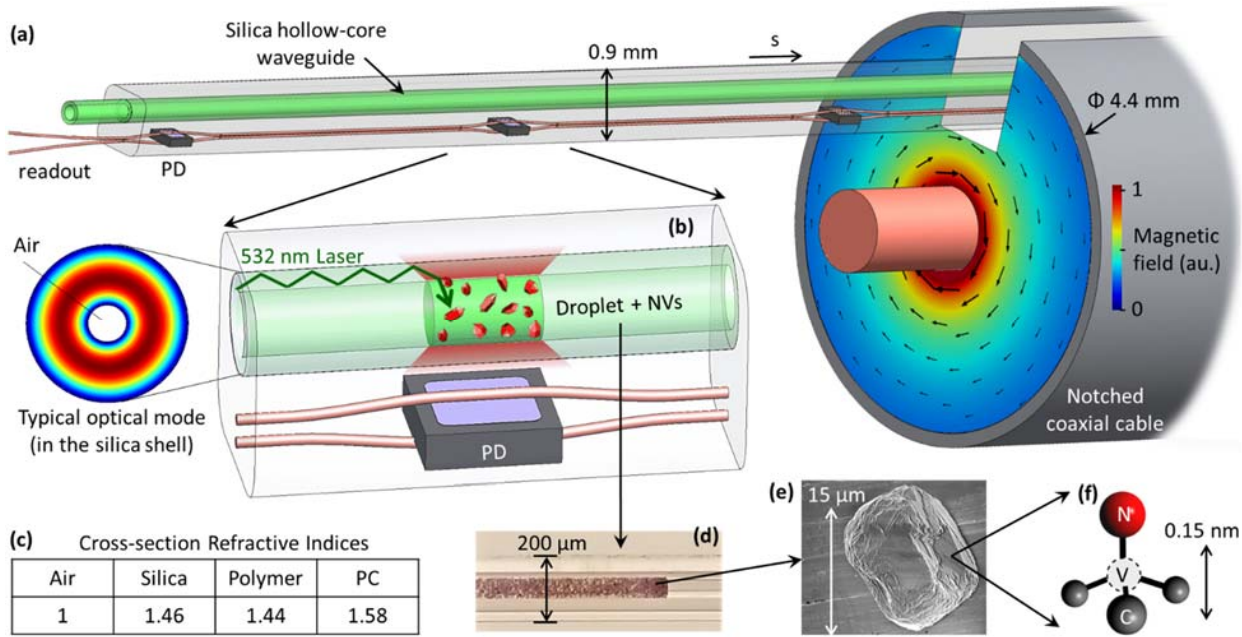

**Figure 2: Fiber magnetometer mechanism. (a)** Schematic of the microfluidic system that allows distributed long-range magnetometry. The hollow-core silica fiber is coated with a polymer that functions as cladding for light guidance. **(b)** Zoom-in on one sensing module. The system is composed of a 532 nm laser source coupled to a hollow-core silica waveguide. Typical donut-shaped optical mode that propagates in the silica shell is shown. We control the droplet position using an air pressure pump connected at the end of the fiber. We align a liquid droplet mixed with >4 ppm NV concentration (estimated by manufacturer) micro-diamond particles (red crystals) with an embedded diode to collect red fluorescence. A notched coaxial cable waveguides the required microwave frequency to control the NV's spin state to provide ODMR. **(c)** Refractive indices table of the cross-section of the fiber. **(d)** Photograph of the NV droplet. **(e)** SEM photograph of a single micro-diamond. **(f) S**implified atomic structure of the NV, where the dashed circle is the vacancy.

A key component of NV spin-based magnetometry is the spin dependence of the NV's fluorescence. Figure 3a shows the NV spin triplet fine structure, which is the origin of the NV spin-dependent fluorescence. The frequency difference between $|m_s = 0\rangle$ and $|m_s = \pm 1\rangle$ magnetic sublevels is given by $D_{gs} \pm 2\gamma B_z$, where $D_{gs} \approx 2.87$ GHz is the NV zero-field splitting, $\gamma \approx 28$ GHz T$^{-1}$ is the NV gyromagnetic ratio, and $B_z$ is the applied magnetic field along the NV axis. The NV fluorescence intensity depends on its spin: it is "bright" in the $|m_s = 0\rangle$ state due to radiative cycling transition and "dark" in the $|m_s = \pm 1\rangle$ due to channeling into a metastable singlet state, which ultimately decays into the $|m_s = 0\rangle$ sublevel. A microwave field moves the ground state population between the $|m_s = 0\rangle$ and $|m_s = \pm 1\rangle$ levels when resonant. This ODMR spectrum[28,29] is shown for a single micro-diamond with zero magnetic field in Figure 3b. A resonance is observed at $D_{gs}$; the contrast, $C$, represents the change in normalized fluorescence intensity with (black) and without (grey) applied MW field. To determine the magnetic field, we modulate (at 6 kHz, 100% depth) the amplitude of the MW field and monitor the NV fluorescence using a lock-in amplifier. The measured lock-in signal is proportional to $C$.

The micro-diamond ensemble (photographed in Figure 3d-inset) within our droplet has multiple NV orientations. Each NV orientation experiences different $B_z$ and thus has different $|m_s = \pm 1\rangle$ resonant frequencies. This diversity of resonant frequencies causes $C$ to decrease as a function $|\mathbf{B}|$ and be insensitive to the direction of $\mathbf{B}$.[30] The NV fluorescence gave a current of about 5 nA for ambient magnetic field which was fed through a 10 MOhm TIA. Monitoring the lock-in signal at $D_{gs}$ (dashed black line) gives a mapping between the lock-in signal and $|\mathbf{B}|$, see Figure 3c. Figure 3d shows the lock-in signal as a function of magnetic field.

To showcase the distributed quantum magnetometry capabilities of our device, we deploy it down a 70 m corridor (Figure 3e). We measured the magnetic field at 102 points along the corridor.

We compare our measurement to a hand-held magnetometer (Alpha Labs Gaussmeter model VGM) as seen in Figure 3f. We observe a spike in the magnetic field of ~180 μT at the location of Airgas cylinders. Figure 3f-inset shows the measured MW loss as a function of MW frequency for the cable assembly. We measure a loss of 49.5 dB over 90 m, which extrapolates to ~0.5 dB/m at 3 GHz. In principle, the MW loss could be further reduced by using better coaxial cable at the expense of the diameter of the cable system.

We use the mapping in Figure 3c to measure a 40 s period square wave applied with an external electromagnet, see Figure 3g. The measured noise is 55 nT, which extrapolates to a sensitivity of $63\pm 5$ nT/$\sqrt{\text{Hz}}$ when accounting for the 100 ms time constant and the 0.78 Hz equivalent noise bandwidth of the lock-in amplifier (24 dB/oct roll-off, see SI). The PDs other than the one performing the sensing are not exposed to either the guided light nor ambient light as the coaxial cable is opaque. The dark current from all the parallel-connected PDs is below the shot noise from the sensing PD.

Our fiber-integrated magnetometer is sealed within a waterproof cladding, making it resilient under adverse conditions. Figure 4a is a photograph of the cable assembly placed into a water tank. A magnet was placed in the water and moved horizontally using a micrometer. The lock-in signal as a function of MW frequency for different magnetic fields is shown in Figure 4b. After three days of measurements within the water tank, we did not notice a degradation in the fiber magnetometry performance (see Figure 4b-inset). We notice that moving the droplet in and out from the PD location and performing repeated measurements yields a systematic error that equals the sensitivity of our magnetometer.

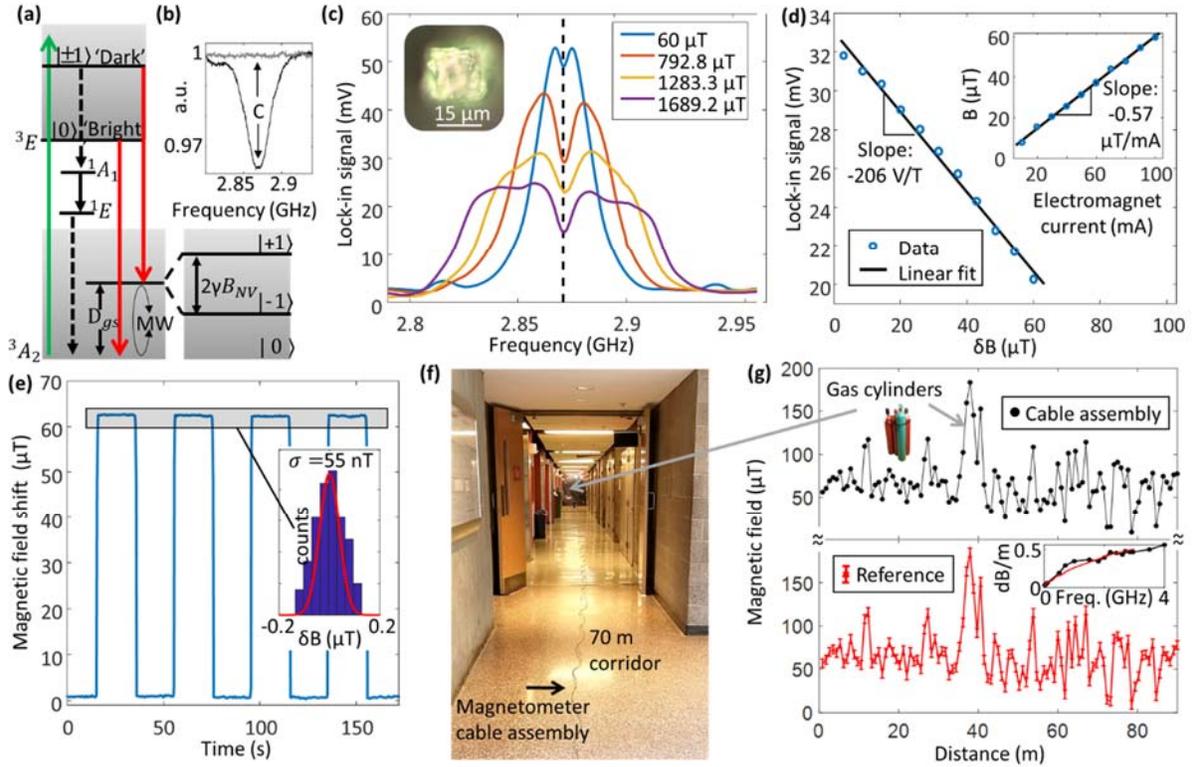

**Figure 3: ODMR spectra and level structure**. **(a)** Energy level diagram of the NV center. Magnetic fields are determined by measuring the NV resonance frequency for the $|m_s = \pm 1 >$ states with radiative (solid lines) and non-radiative (dashed lines) transitions shown. **(b)** Measured ODMR spectrum of a micro-diamond with (black) and without (grey) MW fields applied taken using a microscope. C represents the change in measured normalized intensity due to the application of MW fields. **(c)** Lock-in output of NV micro-diamond ensemble as a function of magnetic field, determined by a fit to the theoretical model (fit not shown).[30] The left inset is a micrograph of a single micro-diamond. **(d)** We place an electromagnet near the cable and we measure the lock-in signal from the fiber magnetometer while changing the electromagnet magnetic field, calibrated in the inset. **(d-inset)** We sweep the electromagnet current while measuring the magnetic field. **(e)** We measure the magnetic field applied by an external electromagnet by monitoring the lock-in signal at $D_{gs}$ (dashed line in c). The applied magnetic

field is a 40 s period square wave with an amplitude of ~60 µT. The measured standard deviation is 55 nT. The histogram of the deviations from the mean is shown in the inset. **(f)** We deploy the cable assembly in a 70 m corridor, and **(g)**, measure the magnetic field magnitude at every meter. We compare the quantum fiber magnetometry results (black dots) to a hand-held magnetometer (red error bars). The quantum fiber magnetometry error bars are smaller than the point's shown. We measure a peak in the magnetic field at the location of the gas cylinders. **(g-inset)** We measure the MW loss of the notched coax (black) and compare it to the loss of the untreated cable as reported by the manufacturer (red).

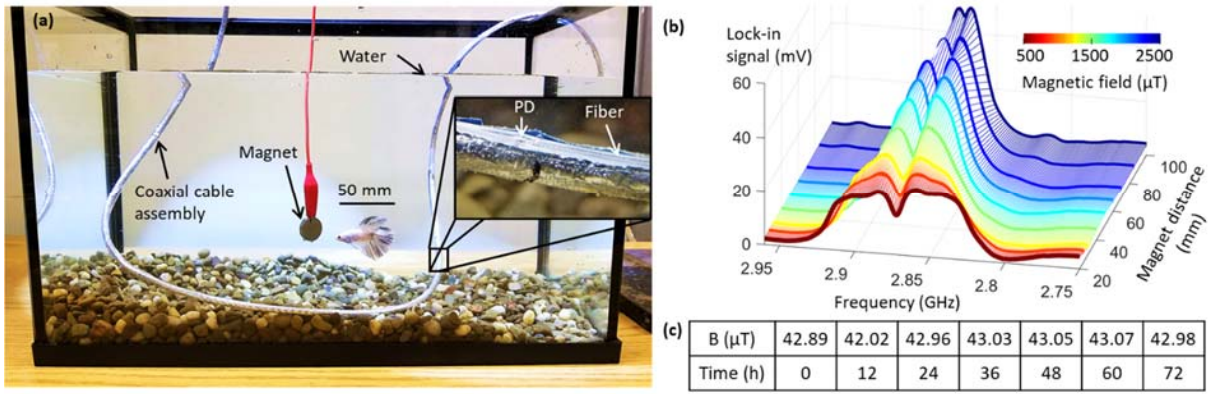

**Figure 4: Underwater Magnetometry. (a)** Photograph of the experiment. We insert our cable magnetometer into a water tank where its waterproof cladding protects it. Inset - zoom in on the notched cable where the PD is visible, and a diamond droplet is located. **(b)** Lock-in signal as a function of MW frequency at different magnet distances. By moving the magnetic horizontally, we change the magnetic field distance. The color represents the magnetic field values determined from the ODMR curves. (c) We remove the magnet and perform repeated measurements over a 3-

day period.

---

Finally, we use the unique distributed nature of our fiber magnetometer to localize and quantify a magnetic source in space. Figure 5a shows the experimental schematic where a magnet cube ($3 \times 3 \times 3$ mm) is placed 1.3 cm away from the fiber. To demonstrate the capability to localize a magnetic object spatially, we align the fiber so that adjacent sensing modules are 3, 2.2, 5.7, and 3 cm apart, respectively, by making loops in the fiber. Figure. 5b shows the measured lock-in signal as a function of MW frequency at each of the sensing modules. The fit (grey) of the spectra determines the applied magnetic field at each node, giving the magnetic field distribution along the fiber. Modeling the applied magnetic field as a magnetic dipole, allows us to determine the location, orientation, and magnitude of the magnetic object.[30] For example, we determine the axial and radial position with an accuracy of 1.87 and 2.87 mm, respectively. We compare the numeric solution (red) to a direct measurement (black) in Figure 5c. The magnetic field magnitude along the fiber magnetometer is relatively insensitive to the magnet orientation; therefore, we observe error bars of ~10 degrees for the orientation parameters $\theta$ and $\beta$.

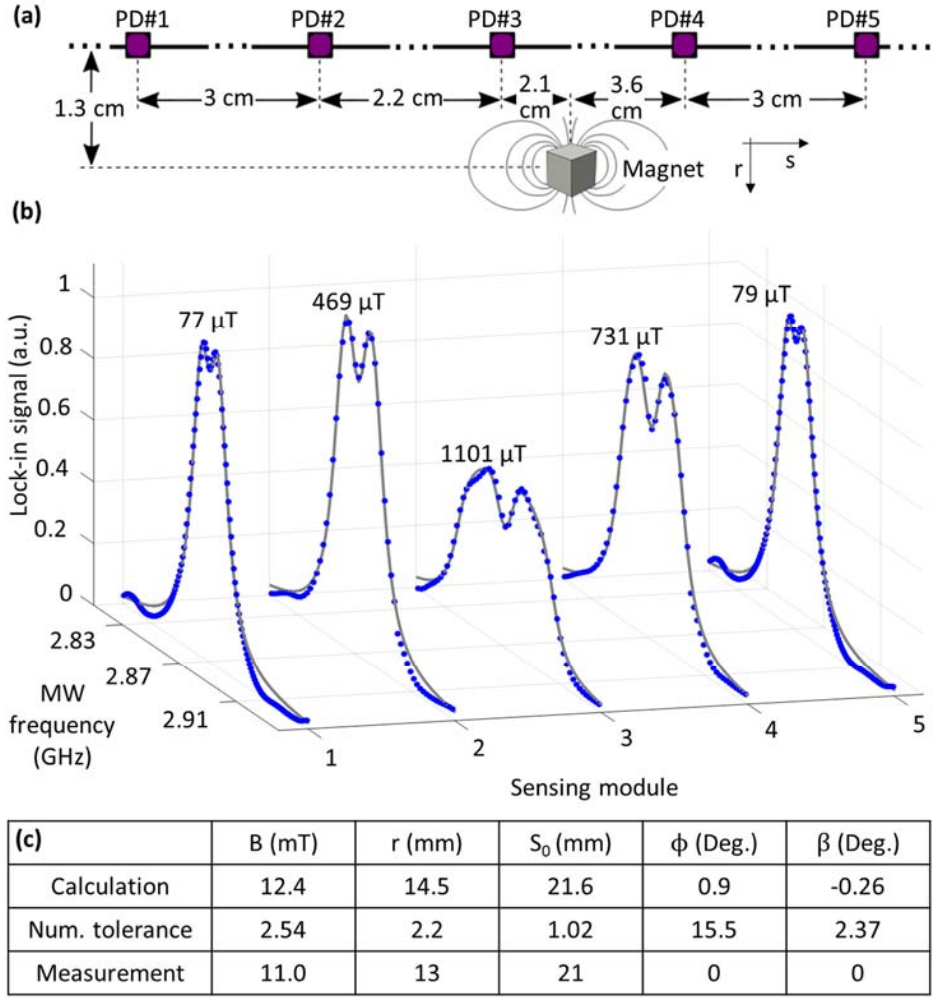

| (c) | B (mT) | r (mm) | $S_0$ (mm) | φ (Deg.) | β (Deg.) |
|---|---|---|---|---|---|
| Calculation | 12.4 | 14.5 | 21.6 | 0.9 | -0.26 |
| Num. tolerance | 2.54 | 2.2 | 1.02 | 15.5 | 2.37 |
| Measurement | 11.0 | 13 | 21 | 0 | 0 |

**Figure 5, Experimental results, localizing external magnet. (a)** We place a magnet at a radial distance of 1.3 cm from the fiber. The closest photodiode is 2.47 cm away from the magnet. **(b)** ODMR spectra as a function of *s*. We fit the data (blue) to a theoretical model (gray) to determine the magnetic field magnitude at each sensing module (appears above each fit). **(c)** Table containing the numerical solution to the magnetic field magnitude, orientation, and location using the five measurements of |**B(s)**| compared to a direct measurement.

Moving the droplet axially within the fiber in a reversible manner is done at typical speeds between 10-30 mm/s due to the hollow-silica dimensions and the oil viscosity, which is comparable to similar microfluidic optical fibers.[31] The scanning time can be decreased by increasing the diameter of the fiber at the cost of collection efficiency, as the diode droplet distance increases. Another approach to improve the scanning speed is by incorporating multiple droplets within the hollow fiber, where only one droplet is aligned with a sensing site. The average spacing between photodiodes is 17 cm, allowing up to 170 droplets simultaneously within our device for a droplet size of 1 mm. In each measurement instance, only one droplet will be aligned with a photodiode. Moving the 170 droplets at a speed of ~10 mm/s allows total scanning time of less than 10 minutes across a 1 km cable. However, there is a tradeoff between scanning speed and sensitivity: when working with multiple droplets, it is required to decrease their size and limit their amount due to optical losses from strong scattering in each droplet location (we measure a loss of 2.4 dB per 1 mm droplet) this means the sensitivity will be degraded due to optical loss. To achieve simultaneous sensing from multiple detection sites using the presented technique, we propose to modify the PD chip and to include a modulator in them. By that, each PD will transmit its measurement in a different modulation frequency, and the separation can be made using a Lock-in. Further improvements can be made by integrating custom CMOS-NV quantum sensing chips for localized spin control and readout.[32]

Several improvements of the sensitivity of this fiber magnetometer can be made. As shown in Figure S4, the dominant noise source in our experiment is shot noise from the pump laser. We estimate that an additional order of magnitude improvement in sensitivity is possible by improving green filtering by ~20 dB. Two possible solutions to filter the green light are inserting a thin free-space optical filter within the cladding or modifying the cladding to have a photonic Bragg grating layer [31]. The noise would then be dominated by the shot noise from the NV centers (see the purple line in Figure S4). The SNR of this sensor could also be improved by increasing the total amount of collected NV fluorescence as shown in bulk diamond NV ensemble magnetometers, which have recently demonstrated a sensitivity of 290 pT/$\sqrt{\text{Hz}}$ for DC magnetometry[33]. One can achieve additional improvement by utilizing NV centers with improved spin coherence. These NVs would have steeper slopes than the one shown in Figure 3c, at the expense of dynamic range. Such NV centers are found in high-quality nanodiamonds[34], within a CVD bulk diamond[14], or an isotopically purified diamond[21].

Several attempts have been made to avoid the need for microwave fields in NV-based magnetometry[35–37]. However, these techniques require both large bias magnetic field at each node and the use of single crystal diamond where the NV's axis is aligned with the bias field. Other all-optical approaches such as coherent population trapping[38] or electromagnetically induced transparency[39] appear to be only compatible with a cryogenic operation. For those reasons, we chose to integrate the RF transmission line in the shape of coaxial cable throughout our fiber.

In conclusion, we have demonstrated a distributed fully-integrated magnetic field sensor that enables localization and quantification of magnetic fields along a hundred meter-scale length with a sensitivity of 63 nT$\sqrt{\text{Hz}}$ and a spatial resolution of 17 cm. Such a distributed water-immersible quantum fiber magnetometer promises new applications for remote detection and tracking in a range of fields including geophysics, ferrous metal detection, and biomedical sensing.

### Acknowledgments

This work was supported in part by the Government of Israel, Ministry of Defense, Mission to the USA No. 4440656472, and in part by MIT Materials Research Science and Engineering Center (MRSEC) through the MRSEC Program of the National Science Foundation under award number

DMR-1419807. This work was also supported in part by the US Army Research Laboratory and the US Army Research Office through the Institute for Soldier Nanotechnologies, under contract number W911NF-13-D-0001, with funding provided by the Air Force Medical Services. This work was also supported by the Assistant Secretary of Defense for Research and Engineering under Air Force Contract numbers FA8721-05-C-0002 and FA8702-15-D-0001. C.F. acknowledges support from Master Dynamic Limited, and by the Army Research Office Multidisciplinary University Research Initiative (ARO MURI) biological transduction program. The authors express their gratitude to David Bono, Collin Marcus, and Mohamed Ibrahim for their advice on the TIA and lock-in amplifier measurements; Michael Tarkanian for his help on the cable machining, and Michael Rein for his advice and support.

**Author contributions**

S.M. fabricated the devices and performed the experiments. S.M. and C.F. analyzed the data. D.R.E. and Y.F. supervised the research. All authors contributed to writing the manuscript.

**Additional Information**

The authors declare no competing financial and non-financial interests.